\documentclass[12pt]{article}
\usepackage{graphics}
\usepackage{enumerate}
\usepackage{amssymb,epsfig,amsmath,euscript,array}
\usepackage{axodraw}
\usepackage{subfigure}
\usepackage{color}
\usepackage[sort&compress,square]{natbib}
\makeatletter
\@addtoreset{equation}{section}
\makeatother

\definecolor{blue}{rgb}{0,0,1}

\definecolor{red}{rgb}{1,0,0}

\definecolor{green}{rgb}{0,1,0}

\newcounter{multieqs}

\newcommand{\be}{\begin{equation}}
\newcommand{\ee}{\end{equation}}

\newcommand{\bm}[1]{\mbox{\boldmath $#1$}}

\def\bd{\begin{document}}
\def\ed{\end{document}}
\def\nn{\nonumber}
\def\bea{\begin{eqnarray}}
\def\eea{\end{eqnarray}}
\let\bm=\bibitem
\let\la=\label

\newcommand{\EQ}[1]{\begin{equation} #1 \end{equation}}
\newcommand{\AL}[1]{\begin{subequations}\begin{align} #1 \end{align}\end{subequations}}
\newcommand{\SP}[1]{\begin{equation}\begin{split} #1 \end{split}\end{equation}}
\newcommand{\ALAT}[2]{\begin{subequations}\begin{alignat}{#1} #2 \end{alignat}\end{subequations}}
\def\beqa{\begin{eqnarray}}
\def\eeqa{\end{eqnarray}}
\def\beq{\begin{equation}}
\def\eeq{\end{equation}}

\def\hf{{\textstyle \frac{1}{2}}}
\def\wbar{\bar w}
\def\mubar{\bar\mu}
\def\abar{\bar a}
\def\sigmabar{\bar\sigma}
\def\etabar{\bar\eta}
\def\zetabar{\bar\zeta}
\def\mubar{\bar\mu}
\def\nubar{\bar\nu}
\def\N{{\cal N}}
\def\sst{\scriptscriptstyle}
\def\thetabar{\bar\theta}
\def\Tr{{\rm Tr}}
\def\one{\mbox{1 \kern-.59em {\rm l}}}
 \def\Nh{\hat{N}}

\newlength{\myVSpace}% the height of the box
\newcommand\xstrut{\raisebox{-.5\myVSpace}% symmetric behaviour,
  {\rule{0pt}{\myVSpace}}%
}

\def\a{\alpha}      \def\da{{\dot\alpha}}
\def\b{\beta}       \def\db{{\dot\beta}}
\def\c{\gamma}  \def\G{\Gamma}  \def\cdt{\dot\gamma}
\def\d{\delta}  \def\D{\Delta}  \def\ddt{\dot\delta}
\def\e{\epsilon}        \def\vare{\varepsilon}
\def\f{\phi}    \def\F{\Phi}    \def\vvf{\f}
\def\h{\eta}
\def\k{\kappa}
\def\l{\lambda} \def\L{\Lambda}
\def\m{\mu} \def\n{\nu}
\def\o{\omega}
\def\p{\pi} \def\P{\Pi}
\def\r{\rho}
\def\s{\sigma}  \def\S{\Sigma}
\def\t{\tau}
\def\th{\theta} \def\Th{\Theta} \def\vth{\vartheta}
\def\X{\Xeta}
\def\z{\zeta}

\def\cA{{\cal A}} \def\cB{{\cal B}} \def\cC{{\cal C}}
\def\cD{{\cal D}} \def\cE{{\cal E}} \def\cF{{\cal F}}
\def\cG{{\cal G}} \def\cH{{\cal H}} \def\cI{{\cal I}}
\def\cJ{{\cal J}} \def\cK{{\cal K}} \def\cL{{\cal L}}
\def\cM{{\cal M}} \def\cN{{\cal N}} \def\cO{{\cal O}}
\def\cP{{\cal P}} \def\cQ{{\cal Q}} \def\cR{{\cal R}}
\def\cS{{\cal S}} \def\cT{{\cal T}} \def\cU{{\cal U}}
\def\cV{{\cal V}} \def\cW{{\cal W}} \def\cX{{\cal X}}
\def\cY{{\cal Y}} \def\cZ{{\cal Z}}

\def\ua{\underline{\alpha}}
\def\ub{\underline{\phantom{\alpha}}\!\!\!\beta}
\def\uc{\underline{\phantom{\alpha}}\!\!\!\gamma}
\def\um{\underline{\mu}}
\def\ud{\underline\delta}
\def\ue{\underline\epsilon}
\def\una{\underline a}\def\unA{\underline A}
\def\unb{\underline b}\def\unB{\underline B}
\def\unc{\underline c}\def\unC{\underline C}
\def\und{\underline d}\def\unD{\underline D}
\def\une{\underline e}\def\unE{\underline E}
\def\unf{\underline{\phantom{e}}\!\!\!\! f}\def\unF{\underline F}
\def\unm{\underline m}\def\unM{\underline M}
\def\unn{\underline n}\def\unN{\underline N}
\def\unp{\underline{\phantom{a}}\!\!\! p}\def\unP{\underline P}
\def\unq{\underline{\phantom{a}}\!\!\! q}
\def\unQ{\underline{\phantom{A}}\!\!\!\! Q}
\def\unH{\underline{H}}

\def\As {{A \hspace{-6.4pt} \slash}\;}
\def\bs {{b \hspace{-6.4pt} \slash}\;}
\def\Ds {{D \hspace{-6.4pt} \slash}\;}
\def\ds {{\del \hspace{-6.4pt} \slash}\;}
\def\ss {{\s \hspace{-6.4pt} \slash}\;}
\def\ks {{ k \hspace{-6.4pt} \slash}\;}
\def\ps {{p \hspace{-6.4pt} \slash}\;}
\def\pas {{{p_1} \hspace{-6.4pt} \slash}\;}
\def\pbs {{{p_2} \hspace{-6.4pt} \slash}\;}

\def\Fh{\hat{F}}
\def\Vh{\hat{V}}
\def\Xh{\hat{X}}
\def\ah{\hat{a}}
\def\xh{\hat{x}}
\def\yh{\hat{y}}
\def\ph{\hat{p}}
\def\xih{\hat{\xi}}

\def\psit{\tilde{\psi}}
\def\Psit{\tilde{\Psi}}
\def\tht{\tilde{\th}}

\def\At{\tilde{A}}
\def\Qt{\tilde{Q}}
\def\Rt{\tilde{R}}
\def\Nt{\tilde{N}}

\def\at{\tilde{a}}
\def\st{\tilde{s}}
\def\ft{\tilde{f}}
\def\pt{\tilde{p}}
\def\qt{\tilde{q}}
\def\vt{\tilde{v}}
\def\nt{\tilde{n}}

\def\delb{\bar{\partial}}
\def\bz{\bar{z}}
\def\bD{\bar{D}}
\def\bB{\bar{B}}

\def\bk{{\bf k}}
\def\bl{{\bf l}}
\def\bp{{\bf p}}
\def\bq{{\bf q}}
\def\br{{\bf r}}
\def\bx{{\bf x}}
\def\by{{\bf y}}
\def\bR{{\bf R}}
\def\bV{{\bf V}}

\def\d{\delta}\def\D{\Delta}\def\ddt{\dot\delta}

\def\pa{\partial} \def\del{\partial}
\def\xx{\times}
\def\uno{\mbox{1 \kern-.59em {\rm l}}}

\def\trp{^{\top}}
\def\inv{^{-1}}
\def\dag{{^{\dagger}}}
\def\pr{^{\prime}}

\def\rar{\rightarrow}
\def\lar{\leftarrow}
\def\lrar{\leftrightarrow}

\newcommand{\0}{\,\!}      %this is just NOTHING!
\def\one{1\!\!1\,\,}
\def\im{\imath}
\def\jm{\jmath}

\newcommand{\tr}{\mbox{tr}}
\newcommand{\slsh}[1]{/ \!\!\!\! #1}

\def\vac{|0\rangle}
\def\lvac{\langle 0|}

\def\hlf{\frac{1}{2}}
\def\ove#1{\frac{1}{#1}}

\def\Box{\square}
\def\ZZ{\mathbb{Z}}
\def\CC#1{({\bf #1})}
\def\bcomment#1{}
\def\bfhat#1{{\bf \hat{#1}}}
\def\VEV#1{\left\langle #1\right\rangle}
\def\vev#1{\langle{#1}\rangle}

\newcommand{\ex}[1]{{\rm e}^{#1}} \def\ii{{\rm i}}

\def\rr{{\rm r}} \def\rs{{\rm s}}\def\rv{{\rm v}}
\def\ri{{\rm i}}\def\rj{{\rm j}}

\newcommand{\lrbrk}[1]{\left(#1\right)}
\newcommand{\sfrac}[2]{{\textstyle\frac{#1}{#2}}}

\font\mybb=msbm10 at 12pt
\def\bb#1{\hbox{\mybb#1}}

\font\myBB=msbm10 at 18pt
\def\BB#1{\hbox{\myBB#1}}

\begin{document}
\noindent
\hspace*{13.5cm}DESY 05-230\\
\hspace*{13.5cm}IPPP/05/76\\
\hspace*{13.5cm}DCPT/05/146

\vspace{25pt}

\begin{center}

{\Large \bf Noncommutativity, Extra Dimensions,\\
 and Power Law Running
in the Infrared 
\\[1.5ex]
}

\vspace{30pt}

{\bf Steven A. Abel$^1$, Joerg Jaeckel$^2$, Valentin V.  Khoze$^3$ and Andreas Ringwald$^2$}

{\small \em
{}$^1$Department of Mathematical Sciences, University of Durham,
Durham, DH1 3LE, UK\\
{}$^2$Deutsches Elektronen-Synchrotron DESY,
Notkestrasse 85, D-22607  Hamburg, Germany\\
{}$^3$Department of Physics, University of Durham,
Durham, DH1 3LE, UK\\

\vspace{10pt}

{\sffamily \tt
s.a.abel@durham.ac.uk, joerg.jaeckel@desy.de, valya.khoze@durham.ac.uk, andreas.ringwald@desy.de}
}

\vspace{30pt}
{\bf Abstract}
\end{center}
We investigate the running gauge couplings of U($N$) noncommutative gauge theories with compact
extra dimensions.
Power law running of the trace-U(1) gauge coupling in the ultraviolet
is communicated to the infrared by ultraviolet/infrared mixing, whereas the
SU($N$) factors run exactly as in the commutative theory. 
This results in theories where the experimentally excluded trace-U(1) factors 
decouple with a power law running of the momentum in the extreme infrared, 
effectively hiding them from detection.

\noindent {}
\setcounter{page}{0} \thispagestyle{empty}

\newpage

\section{Introduction}
Gauge theories on spaces with noncommuting coordinates,
\begin{equation}
[x^\mu,x^\nu]=i\,\theta^{\mu\nu} \ ,
\end{equation}
are an interesting
class of quantum field theories with intriguing and sometimes unexpected features.
These noncommutative models can arise naturally as low-energy effective theories from string
theory and D-branes \cite{Connes:1997cr,Douglas:1997fm,Chu:1998qz,Seiberg:1999vs}.
As field theories they must satisfy a number of restrictive
constraints, and this makes them particularly interesting and challenging
for the purposes of particle physics model building.
For general reviews of noncommutative gauge theories the reader can consult
Refs.~\cite{Seiberg:1999vs,Douglas:2001ba,Szabo:2001kg}.

In the context of noncommutative Standard
Model building, there is a number of generic features
which are very constraining~\cite{Khoze:2004zc}. Particularly important
for the present discussion is the fact that
the gauge groups are restricted to be U($N$)~\cite{Matsubara:2000gr,Armoni:2000xr} or products thereof, and
that fields can transform only in (anti-)fundamental, bi-fundamental
or adjoint representations~\cite{Gracia-Bondia:2000pz,Terashima:2000xq,Chaichian:2001mu}.

In a recent paper \cite{Jaeckel:2005wt} it was shown that
the consequences for noncommutative gauge theories in four
dimensions are potentially fatal; we would either
observe additional massless degrees of freedom in experiment or the photon would
acquire a Lorentz symmetry violating mass of the order of the supersymmetry
breaking scale, neither effect of which has been seen.

There is one particularly striking feature of noncommutative gauge theories
that superficially looks as if it may solve the problem but in its purely 4-dimensional form does not:
there is a mixing of
ultraviolet (UV) and infrared (IR) effects~\cite{Minwalla:1999px,Matusis:2000jf}
and a consequent asymptotic decoupling of undesirable U(1) degrees of
freedom~\cite{Khoze:2000sy,Hollowood:2001ng,Khoze:2004zc} in the infrared. Unfortunately
the decoupling is logarithmically slow. Indeed the couplings of the SU($N$) and the trace-U(1)
parts of the U($N$) gauge interaction are related as
\begin{equation}
\label{ration1}
\frac{g^2_{\rm U(1)}}{g^2_{\textrm{SU}(N)}}\sim \frac{\log\left(\frac{k^2}
{\Lambda^{2}}\right)}{\log\left(\frac{M^4_{\textrm{NC}}}{\Lambda^{2}k^2}\right)}
\gtrsim 10^{-3}\, ,
\end{equation}
where $\Lambda$ is the strong coupling scale of the SU($N$) and $k^2 \gg \Lambda^2$ is the momentum
scale of a scattering experiment.
The logarithmic dependence on $M_{\textrm{NC}}\sim |\theta|^{-1/2}$
leads to an incredibly tight bound,
$M_{\textrm{NC}} \gg M_{\textrm{\rm Planck}}$.
This makes
such models very unnatural \cite{Jaeckel:2005wt}.

Logarithmic running of couplings is a typical property of four-dimensional field theories. But in
higher dimensions running generically follows a power-law. Thus the question naturally arises as to
whether power law running transfers to the infrared via UV/IR mixing, and whether this is enough to
decouple unwanted U(1) factors. In this paper we show that the answer to both questions is
positive and discuss the bounds.

The rapid power-law decoupling of the trace-U(1) factors opens up possibilities
for particle physics model building based on noncommutative extra-dimensional theories in
the UV. One such approach for embedding the Standard Model into a noncommutative theory
was discussed in \cite{Khoze:2004zc}. Adding compact extra dimensions to these models
can now essentially remove the ubiquitous trace-U(1) degrees of freedom
from the low-energy effective theory.

\section{Power law running in the UV and IR}\label{powerlaw}

We begin by discussing generalities of two-point functions in
noncommutative theories. Consider the polarisation tensor, given by \cite{Matusis:2000jf}
\begin{equation}
\label{poltensor}
\Pi_{\mu\nu}^{AB} = \Pi_1^{AB}(k^2,\tilde k^2) \, \left( k^2 g_{\mu\nu} - k_\mu k_\nu \right)
+ \Pi_{2}^{AB} (k^2, \tilde k^2)\, \frac{\tilde{k}_{\mu}\tilde{k}_{\nu}}{\tilde{k}^2}
\,,
\end{equation}
with
\begin{equation}
\label{ktilde}
\tilde{k}_\mu := \theta_{\mu\nu} k^\nu \,.
\end{equation}
$A$ and $B$ are adjoint indices of U($N$), such that $A,B=0,\ldots N$ and $A,B=0$ select
the trace-U(1) and $A,B=1,\ldots N$ correspond to the SU($N$) parts.
$\Pi_{1}^{AB}$ is directly related to the gauge coupling matrix\footnote{The generators are normalised such that $\textrm{Tr}(t^{A}t^{B})=\frac{1}{2}\delta^{AB}$, as it is standard for SU($N$). For the trace-U(1) coupling
this results in a factor of two compared to the standard U(1) conventions.}
\begin{equation}
\label{defcoupling}
\left(\frac{1}{g^{2}(k)}\right)^{AB}=\left(\frac{1}{g^{2}_{0}}\right)^{AB}+\Pi^{AB}_{1}(k),
\end{equation}
and it is this term we need to evaluate in order to discuss the running
\cite{Khoze:2000sy,Hollowood:2001ng}.
The additional term $\sim \Pi_{2}$ is Lorentz symmetry violating. In supersymmetric theories it is
absent \cite{Matusis:2000jf} and we defer its discussion until Sect. \ref{massec}.

In most of the discussion we will adopt a four-dimensional point of view
in describing extra-dimensional theories. That is, because we are interested
in renormalisation group effects associated with the 4-dimensional momentum, it makes more sense to
include the effects of extra dimensions by considering the effect of a simple Kaluza-Klein tower of states.
(In the UV-complete string models there are other effects which, at one-loop order and in compact dimensions
significantly larger than the string length, will be secondary.)

Intuitively it is obvious that the main factor affecting the running of the gauge couplings will be the
noncommutativity parameter $\tilde{k}$, and in particular how it mixes the additional (compact)
dimensions 
with the ordinary four large dimensions.
We will now give a somewhat heuristic presentation of how $\tilde{k}$ affects the running of the gauge couplings.
A more
precise and general calculation is given in Appendix \ref{detailed} and we will just quote
the results from there in the last part of this section.

\subsection{The UV regime}

Let us start by briefly reviewing power law running in the UV at scales well
above the compactification scale. In the UV regime the planar diagram
(cf. Appendix A) dominates the two point function and so
there is no difference to the ordinary commutative case (see \cite{Dienes:1998vg}). 
Because of this
it is sufficient to use an intuitive approach based on thresholds\footnote{A fuller treatment based on
dimensional regularisation is presented in Appendix B. An even better one
is presented in Ref.~\cite{ghilencea}. In those treatments it becomes
evident that higher-dimensional operators appear in the effective action.
These operators are due to a different form of UV/IR mixing from
regions of KK momenta that are zero in some directions and high in others.
These difficulties are absent for the IR regime which is the main point
of interest in the present discussion so we do not dwell on them here.}.

Consider first the most simple case of one compact extra dimension
of size $M^{-1}_{c}$. Neglecting threshold effects
the one loop running of the gauge coupling in four dimensions typically follows
($t=\log(k)$)
\begin{equation}
\frac{\partial}{\partial t}g^2=\sum_{m^{2}_{i}<k^2}c_{i} g^4,
\end{equation}
where the $c_{i}$ are coefficients depending on the spin and representation of the particle $i$. In the sum only
particles with mass $m^{2}_{i}$ smaller than the momentum scale $k^2$ contribute (cf. Appendix \ref{appirrunning}).
This leads to the typical decoupling
of massive modes.
For simplicity, let us now consider a situation where all particles have (approximately) the same mass $m^2\ll M^{2}_{c}$.
We find
\begin{equation}
\frac{\partial}{\partial t}g^2=-b_0 g^4,\quad \textrm{for}\quad m^2\ll k^2\ll M^{2}_{c},
\end{equation}
where we have chosen the sign of the constant $b_0$ such that it is positive when the theory is asymptotically free.
(For example, in ${\cal N}=2$ supersymmetric pure gauge theory $b_0 =  N /(4\pi^2)$ in this notation.)

Above the compactification scale, more precisely at $m^2+M^{2}_{c}<k^2< m^2+4 M^{2}_{c}$,
the first Kaluza-Klein mode gives an identical contribution to the $\beta$-function, and in general one finds
\begin{equation}
\label{flow0}
\frac{\partial}{\partial t}g^2=-N_{\textrm{KK}}(k)b_0 g^4,
\end{equation}
where $N_{\textrm{KK}}(k)$ is the number of Kaluza-Klein modes (including the zero mode)
contributing at the scale $k$. Since the mass of the $n$th Kaluza-Klein mode is given by
$\sqrt{m^2+n^2M^{2}_{c}}$ one easily finds the approximate formula
\begin{equation}
\label{number}
N_{\textrm{KK}}(k)\approx C_{1} \frac{k}{M_{c}} \quad\textrm{for}\quad k\gg M_{c},
\end{equation}
where we have introduced the constant $C_{1}$ to account for the details of the compactification
and threshold effects (cf. also Appendix \ref{detailed}).
This already suggests power law running. More precisely,
one easily checks that for $k^2\gg M^{2}_{c}$ and appropriate initial conditions the solution approaches
\begin{equation}
\label{powersolution}
g^{2} \approx \frac{1}{C_{1} b_0}\frac{M_{c}}{k}
\end{equation}
which is indeed a power law.

Expressions \eqref{number} and \eqref{powersolution} are easily generalized to arbitrary dimension
$D=n+4$ ($k^2\gg M^{2}_{c}$)
\begin{eqnarray}
\label{flow}
\frac{\partial}{\partial t}g^2\!\!&=&\!\!-N_{\textrm{KK}}(k)b_0 g^4,\\\nonumber
 N_{\textrm{KK}}(k)\!\!&\approx&\!\!  C_{n}\left(\frac{k}{M_{c}}\right)^{n} ,\\\nonumber
 g^{2}\!\!&\approx&\!\!\frac{n}{C_{n}b_0}\left(\frac{M_{c}}{k}\right)^{n},
\end{eqnarray}
where again the constant $C_{n}$ depends on the details of
the compactification.

The flow equation \eqref{flow} for the running coupling can be also discussed
using the more natural effective coupling $\hat{g}^2$ of the $D$-dimensional theory,
\begin{equation}
\label{rescaling}
\hat{g}^2=\left(\frac{k}{M_{c}}\right)^{n}g^2.
\end{equation}
From the lower-dimensional viewpoint
\eqref{rescaling} can be understood by remembering that the amplitudes of all Kaluza-Klein modes add up and therefore
increase the effective coupling by a factor $N_{\textrm{KK}}$.
Inserting \eqref{rescaling} into \eqref{flow} yields the flow equation for $\hat{g}^{2}$,
\begin{equation}
\label{rescaled}
\frac{\partial}{\partial t}\hat{g}^2=n\hat{g}^{2}-C_{n}b_{0}\hat{g}^{4}
=(n-C_{n}b_0\hat{g}^{2})\hat{g}^{2}, \quad\textrm{for}\quad k^2\gg M^{2}_{c}.
\end{equation}
If we start at small values for $\hat{g}^2$ the coupling increases toward the infrared until it reaches a fixed
point at $\hat{g}^2_{\rm fixed}=\frac{n}{C_{n}b_0}$.
The corresponding coupling of the 4-dimensional theory is then
\begin{equation}
\label{fixed4d}
g^2_{\rm fixed} (k) =\hat{g}^2_{\rm fixed} \left(\frac{M_{c}}{k}\right)^{n}=
\frac{n}{C_{n}b_0}\left(\frac{M_{c}}{k}\right)^{n},
\end{equation}
in agreement with the last equation in \eqref{flow}.
This discussion implies that power-law running in extra dimensions originates
from a fixed point in the effective higher-dimensional coupling constant $\hat{g}^2$.
This implies that the power-law running of $g^2$ is a strong coupling phenomenon in terms
of $\hat{g}^2$
and one should exercise caution since Eqs. \eqref{flow} and \eqref{rescaled} are one-loop results.
In particular a large number of extra-dimensions increases the value of the fixed point coupling
and the approximation may break down. The issues of existence of a fixed point
of $\hat{g}^2$ were investigated in literature on extra-dimensional
gauge theories, see e.g. \cite{Gies:2003ic}.

From now on we will continue assuming that (ordinary commutative) extra-dimensional
gauge theories do provide a power-law running of the coupling in the extreme ultraviolet
(i.e. at energies well above the compactification scale).
We will then show that in noncommutative settings the mixing between ultraviolet and infrared
degrees of freedom will induce in the extreme infrared a power-law decoupling of the trace-U(1) degrees
of freedom.

\subsection{The UV/IR mixing}\label{uvirmix}

A novel feature of quantum field theories formulated on noncommutative spaces
is that Wilsonian universality can be violated by the UV/IR mixing effects.
As explained in Refs.~\cite{Minwalla:1999px,Matusis:2000jf} the use of the star product
\begin{equation}
(\phi * \varphi) (x) \equiv \phi(x)\  e^{{i\over 2}\theta^{\mu\nu}
\stackrel{\leftarrow}{\partial_\mu}
\stackrel{\rightarrow}{\partial_\nu}} \  \varphi(x)  \label{stardef}
\end{equation}
in defining noncommutative field theories
mixes up ultraviolet and infrared degrees of freedom such that the high-energy degrees
of freedom do not decouple completely. Instead they can affect degrees of freedom in the extreme
infrared.
The UV/IR mixing in noncommutative theories arises from the fact that due to \eqref{stardef}
certain classes of Feynman diagrams acquire factors of the form
$e^{i k_\mu \theta^{\mu\nu} p_\nu}$
(where $k$ is an external momentum and $p$ is a loop momentum) compared to their commutative
counter-parts.
At large values of the loop momentum $p$,
the oscillations of $e^{i k_\mu \theta^{\mu\nu} p_\nu}$
improve
the convergence of the loop integrals. However, as the external momentum vanishes, $k \to 0,$
the divergence reappears and
what would have been a UV divergence is now reinterpreted as an IR divergence instead.

Essentially,
any effective field theory description of a noncommutative theory
applies at energy (or momentum) scales $k$ in the window,
$\Lambda_{\rm IR}^{\rm induced} < k < \Lambda_{\rm UV}.$
Here $\Lambda_{\rm UV}$ is the UV cut-off
of the effective theory -- which is the scale above which all higher-energy degrees of freedom
of microscopic theory were integrated out. Noncommutativity, and specifically UV/IR mixing effects,
then induce an effective IR cut-off, $\Lambda_{\rm IR}^{\rm induced} \sim M_{\textrm{NC}}^2 /\Lambda_{\rm UV},$
where $M_{\textrm{NC}}$ is the noncommutativity mass-scale, $M_{\textrm{NC}}^2 \sim 1/|\theta|.$
In order to be able to to probe the physics below the IR cut-off $\Lambda_{\rm IR}^{\rm induced}$,
one needs to refine the effective field theory description by raising appropriately the UV cut-off $\Lambda_{\rm UV}.$
Every time physics above $\Lambda_{\rm UV}$ changes, this affects physics below $\Lambda_{\rm IR}^{\rm induced}.$

In a noncommutative U($N$) gauge theory the UV/IR mixing effects contribute to the
$A=0=B$ component of the polarisation tensor
$\Pi_{\mu\nu}^{AB}$ in Eq.~\eqref{poltensor}. This affects the running
coupling $(1/g^2)$
of the trace-U(1) gauge fields via \eqref{defcoupling}.
Our goal is to demonstrate that starting with a 4-dimensional noncommutative effective theory
and embedding it in the UV
into an extra-dimensional gauge theory will induce a rapid power-like decoupling of the unwanted
trace-U(1) degrees of freedom of the original theory in the infrared.

\subsection{IR running -- noncommutativity restricted to 4 dimensions}\label{simple}

As specified in Eq. \eqref{poltensor} $\Pi_{1}$ and therefore the gauge coupling
depends on the additional scale $\tilde{k}$
(cf. \cite{Matusis:2000jf,Minwalla:1999px,Hollowood:2001ng,Khoze:2000sy})
$\tilde{k}^{\mu}=\theta^{\mu\nu}k_{\nu}$.
In fact, the coupling depends only on the absolute values $|\tilde{k}|$ as well as
$|k|$, as can be seen from the discussion in the Appendix \ref{appirrunning}.

Since we are mostly interested in low-energy physics (compared to the compactification scale)
the effects of extra dimensions can contribute only through loops in perturbation theory.
Thus the external momenta $k_\mu$ are taken to be 4-dimensional,
i.e. external particles will not include excited Kaluza-Klein modes, while
internal loop momenta $p_\mu$ (in Feynman diagrams) are kept general.

In this section we consider a scenario where
only the four infinite dimensions are noncommutative,
\begin{equation}
\label{fourdcase}
\theta^{\mu\nu} \neq 0, \quad \theta^{\mu b}=0, \quad
\theta^{a b}=0
\end{equation}
where
$\mu,\nu=0,\ldots,3$ and $a, b=4,\ldots,3+n.$

We can then simplify the discussion of UV/IR mixing by writing
\begin{equation}
\label{absktilde}
|\tilde{k}| =\,  M^{-2}_{\textrm{NC}}|k|,
\end{equation}
where $ M_{\textrm{NC}}$ is the noncommutativity mass-scale.
Heuristically, $M^{-2}_{\textrm{NC}} \sim |\theta|.$
More precisely, for $\theta^{\mu\nu}$ in the canonical basis,
\be
\theta^{\mu\nu} = \left(\begin{matrix} 0 & \theta_1 & 0 & 0 \cr
-\theta_1 & 0 & 0 &0 \cr 0 & 0 & 0 & \theta_2 \cr
0 & 0 & -\theta_2 & 0 \end{matrix}\right)
,
\ee
only when $\theta_1 \simeq \theta_2$ one has
$M^{-2}_{\textrm{NC}} = |\theta|.$ Otherwise the scale
$ M_{\textrm{NC}}$ depends on $k_\mu,$
\be
M^{-2}_{\textrm{NC}} =\,  |\theta_2|\,  \sqrt{1 +
\frac{\theta^{2}_{1}-\theta^{2}_{2}}{\theta^2_2} \frac{k_0^2+k_1^2}{k^2}}.
\ee
It is nevertheless a useful scale.

Following the approach of \cite{Khoze:2000sy,Hollowood:2001ng} we show
in Appendix \ref{appirrunning}
that in a 4-dimensional noncommutative gauge theory with all particles of equal
non-zero mass $m$, the trace-U(1)
couplings runs according to
\begin{equation}
\label{irrun}
\frac{\partial}{\partial t}g^2=b^{\textrm{np}}_{0} g^4
\quad\textrm{for} \quad k^2\ll \min\left(M^{2}_{\textrm{NC}},\frac{M^{4}_{\textrm{NC}}}{m^2}\right).
\end{equation}
Here $b^{\textrm{np}}_{0}$ is a positive number
which is the non-planar contribution to the coefficient $b_0,$
see Eq.~\eqref{definitionsb}.

From discussion of the UV/IR mixing in Appendix \ref{appirrunning} and elsewhere
 one can see that in general noncommutative theory when we lower momentum-scale $k^2$ sufficiently,
even very massive modes start to contribute. This holds for Kaluza-Klein modes, too, as long as
we have noncommutativity only in the four infinite
dimensions according to Eq. \eqref{fourdcase}.
In analogy to \eqref{flow} we find
($k^2\ll\min(M^{2}_{\textrm{NC}},\frac{M^{4}_{\textrm{NC}}}{M^{2}_{c}})$)
\begin{eqnarray}
\label{flowir}
\frac{\partial}{\partial t}g^2\!\!&=&\!\! N^{\textrm{IR}}_{\textrm{KK}}(k)b^{\textrm{np}}_0 g^4,
\\\nonumber
 N^{\textrm{IR}}_{\textrm{KK}}(k)\!\!&\approx&\!\!  C^{\textrm{IR}}_{n}\left(\frac{M^{2}_{\textrm{NC}}}{M_{c}k}\right)^{n} ,
 \\ \nonumber
 g^{2}\!\!&\approx&\!\!\frac{n}{C^{\textrm{IR}}_{n}b^{\textrm{np}}_0}\left(\frac{kM_{c}}{M^{2}_{\textrm{NC}}}\right)^{n}.
\end{eqnarray}
The right hand side of the IR flow equation in \eqref{flowir} has the opposite sign
to that of the UV flow equation \eqref{flow}. This implies that the
trace-U(1) coupling $g^2$ becomes small in the IR and the UV regimes. The enhancement by
the $N^{\textrm{IR}}_{\textrm{KK}}(k)$ factor gives the {\em power-like} decoupling
of these unwanted degrees of freedom from the SU($N$) theory (which is unaffected by the UV/IR mixing effects).

The infrared decoupling of the trace-U(1) can be expressed as follows.
As mentioned earlier, our effective field-theoretical description is valid at energy scales
in the region below the ultraviolet cut-off and above the induced infrared cut-off,
$\Lambda_{\rm IR}^{\rm induced} < k < \Lambda_{\rm UV}.$
Naturally, we assume that $\Lambda_{\rm UV} \gg M_{\textrm{NC}}$ so that
$\Lambda_{\rm IR}^{\rm induced} = M_{\textrm{NC}}^2 /\Lambda_{\rm UV}
\ll M_{\textrm{NC}}.$
The extreme infrared region corresponds to $k$ approaching $\Lambda_{\rm IR}^{\rm induced}.$
The IR-value of the trace-U(1) coupling is then given by the last equation in \eqref{flowir},
\be
g^{2}_{\rm IR} \approx \frac{n}{C^{\textrm{IR}}_{n}b^{\textrm{np}}_0}\left(\frac{M_{c}}{\Lambda_{\rm UV}}\right)^{n}.
\ee
At the same time, the last equation in \eqref{flow} gives the U($N$) coupling in the extreme UV:
\be
g^{2}_{\rm UV} \approx
\frac{n}{C_{n}b_0}\left(\frac{M_{c}}{\Lambda_{\rm UV}}\right)^{n}.
\ee
Therefore, in the IR the trace-U(1) returns to the initial value of the U($N$) gauge coupling in the extreme UV up to a factor of order unity.
Thus another
way of expressing the decoupling scenario is that all the U($N$) gauge
couplings are very small in the UV but that the trace-U(1) part
returns to approximately its UV value (or an even smaller one) whereas the SU($N$) factors attain physically
acceptable large coupling by power law running in the UV.
This conclusion is also in agreement with  the 4-dimensional results discussed in detail
in \cite{Khoze:2000sy}. 

\subsection{IR running for arbitrary noncommutativity}

If the matrix
$\theta^{\mu\nu}$ has nonvanishing entries that mix the ordinary four dimensions with the
extra dimensions we may have a non-vanishing
\begin{equation}
\hat{k}^{a}  =  \theta^{a\nu}k_{\nu}\,\,\,\,\,\,\,(a=4\ldots,3+n).
\end{equation}
In the calculation of the polarisation tensor this
leads to phase factors in the sum over the Kaluza-Klein modes,
\begin{equation}
\sum_{m\in\mathbb{Z}^{n}} e^{i\frac{m}{R}\cdot\hat{k}}
\end{equation}
(in addition to the usual $\theta$-dependent phases
in non-planar contributions).
In this situation it is advantageous to directly perform the sum
over Kaluza-Klein modes in the polarisation tensor.
We have done
this explicitly in Appendix \ref{detailed}.
Here we will quote the result (for an ${\mathcal{N}}=2$ supersymmetric U($N$) theory 
without adjoint matter fields),
\begin{equation}
\label{resultshort}
\Pi_{1}
=const+2\frac{C(\mathbf{G})}{(4\pi)^{2}}(4\pi)^{\frac{n}{2}}\Gamma\left(\frac{n}{2}\right) \prod_{i}R_{i}
\left(|\tilde{k}|^{-n}\right),
\end{equation}
where $R_{i}$ are the compactification radii and $\tilde{k}$ is now the total
noncommutative momentum $\tilde{k}^{M}=\theta^{M\nu}k_\nu $ ($M=0\ldots 3+n$).
This equation is valid for
\begin{equation}
\label{regimelow}
k\ll \min\left(M_{c},\frac{M^{2}_{\textrm{NC}}}{M_{c}}\right).
\end{equation}

The fact that the actual running
is now given by replacing the 4-dimensional components of
$\tilde{k}$ with the total ${\tilde k}$ is not too surprising since
the infrared running comes from very ultraviolet modes, i.e. it involves momenta much higher than the
compactification scale where the theory is effectively higher-dimensional. At these scales there is no
distinction between the ordinary four dimensions and the extra dimensions.

Eq. \eqref{resultshort} has the additional
advantage that it already corresponds to the integrated result. It directly gives $g(k)$ without
the need to solve a differential equation
($R_{i}=1/M_{c}$), 
\begin{equation}
\label{resultu1}
g^{2}_{\textrm{U}(1)}(k)
=\frac{1}{A_{\textrm{U}(1)}+\frac{C^{\textrm{IR}}_{n}b^{\textrm{np}}_{0}}{n}\left(\frac{M^{2}_{\textrm{NC}}}{M_{c}k}\right)^{n}}.
\end{equation}
Here we have fixed,
\begin{eqnarray}
C^{\textrm{IR}}_{n}\!\!&=&\!\! \frac{n}{2}(4\pi)^{\frac{n}{2}}\Gamma\left(\frac{n}{2}\right),
\\\nonumber
b^{\textrm{np}}_{0}\!\!&=&\!\! \frac{4}{(4\pi)^{2}}C(\mathbf{G})%N_{\mathbf{G}}
,
\end{eqnarray}
where we still consider the ${\mathcal{N}}=2$ case and none of the matter fields are in the 
adjoint representation\footnote{A generalisation to an arbitrary number of matter multiplets 
can be easily obtained from the results given in the appendices.}. 
$A_{\textrm{U}(1)}$ is a renormalisation constant determined from the bare coupling and the planar diagrams only.
Therefore in the regime \eqref{regimelow} this constant is connected to the gauge coupling
of the SU($N$)-part (up to logarithmic corrections which we neglected in our approximation)
\begin{equation}
\label{resultsun}
g^{2}_{\textrm{SU}(N)}(k)\approx\frac{1}{A_{\textrm{SU}(N)}}\quad\textrm{with}\quad A_{\textrm{U}(1)}=A_{\textrm{SU}(N)}.
\end{equation}

\subsection{Physics in the $|\tilde k|\rightarrow 0 $ limit}
In section 2.2 we noted that UV/IR mixing forces us to include an
IR cut-off matching the UV one, roughly given by 
\[ |k| > \Lambda_{\textrm{IR}} = \frac{M_{\textrm{NC}}^2}{\Lambda_{\textrm{UV}}}.\]
In order to gain more insight about the nature 
of the physical cut-off it is instructive to consider 
specific physical situations that at first sight seem to lead to difficulties.
For example consider the case where 
time and one space dimension, which we choose to be $x^1$,  are completely commutative.
(This corresponds to choosing $\theta_1=0$ in Eq. (2.16)). 
In that case it seems that if we were to perform an experiment 
with $k_{2}=k_3=0$ we have a dichotomy: since
$\tilde k =0$ we could argue that the trace-U(1) coupling is zero, but on the 
other hand since $\theta_{1\mu}=0$ the 
noncommutativity cannot effect physics in the $x^1$ 
direction and the trace-U(1) 
coupling must therefore run in the same way as in a commutative theory. 
This sort of discontinuity (i.e. physics in the $\theta^{\mu\nu}\rightarrow 0$ limit is
not the same is $\theta^{\mu\nu}=0$ physics) is a familiar aspect of 
UV/IR mixing.

To see that there is no paradox, we need to be more careful in considering the 
regions of validity. We have seen that physics is only well defined within 
certain cut-offs: if $\theta^{\mu\nu}$ vanishes in commutative planes, these will not be isotropic. Indeed if the $ij$ plane has $\theta_{ij}$ then we must
define an induced IR cut-off for each 2-plane (corresponding to applying the 
UV cut-off $\Lambda_{\textrm{UV}}$ in each plane). This results in a region of physical validity 
for the noncommutative field theory, outside which one requires a UV complete theory, 
\[ \Lambda_{\textrm{IR}}^{ij} \sim \frac{1}{|\theta_{ij}|\Lambda_{\textrm{UV}}} < |k| < \Lambda_{\textrm{UV}}.\]
We can now make two observations. Firstly the $\theta^{ij}\rightarrow 
0$ limit does not in fact exist since 
by the above we must always have
$ |\theta_{ij}| > \Lambda^{-2}_{\textrm{UV}} $. Second, even if 
$\theta_{01} =0$, there is no $\tilde{k}\rightarrow 0$ limit either since
we have $|\tilde{k}|> \frac{1}{\Lambda_{\textrm{UV}}} $. This means that in our 
experiment there is an extremely narrow wedge near $\tilde{k}=0$ 
within which the field theory is no longer valid and something else, 
presumably string physics, applies.

We can see this in more detail by repeating the analysis of Appendix B but using 
a UV cut-off rather than dimensional regularization. 
We then find
\begin{equation}
\label{resultnew}
\Pi_{1}
=const+2\frac{C(\mathbf{G})}{(4\pi)^{2}}(4\pi)^{\frac{n}{2}}\Gamma\left(\frac{n}{2}\right)\prod_{i}R_{i}
\left( (|\tilde{k}|^2+\Lambda_{UV}^{-2})\right)^{-{\frac{n}{2}}}.
\end{equation}
Thus for $|\tilde{k}| < \frac{1}{\Lambda_{\textrm{UV}}}$ the IR power-law running 
of the trace-U(1) coupling is simply frozen by the UV cut-off, close to its small UV value. 
It is important to realise that in this sense $\Lambda_{\textrm{IR}}^{\textrm{induced}}$ is not a
separate object from $\Lambda_{\textrm{UV}}$ but it results in fact from the same cut-off being
applied to the same physical modes.
Whatever happens below this cut-off is beyond our control and belongs to the realm of the 
UV complete theory.

Another set-up that is relevant in this context is an extra 
dimensional theory where noncommutativity is confined entirely 
to the extra dimensions. 
Then the $\tilde{k}\approx 0$ wedge is precisely 
where we are living since all our experiments are performed with zero 
extra-dimensional momenta. Thankfully the above discussion means that 
we do not need string theory to describe everyday physics: 
essentially all that is seen in the 4 commutative dimensions 
is commutative physics, with the trace-U(1) gauge coupling remaining 
frozen at its UV value\footnote{This tacitly assumes that the effects of
the UV completion can be qualitatively described by the presence of a cutoff, i.e. the UV completion  merely adds
some threshold effects to the low energy physics.} (of course this may no longer be the case in 
processes involving external Kaluza-Klein states).

\section{Lorentz violating mass term for trace-U(1)}\label{massec}
In noncommutative field theories the gauge coupling is not the only part of the polarisation tensor
that is affected by power law running. Recall that in noncommutative field
theories the (4-dimensional) polarisation tensor has
an additional Lorentz symmetry violating part \cite{Matusis:2000jf,Khoze:2000sy}, which is
called $\Pi_{2}$ in Eq. \eqref{poltensor}.

It is well known
\cite{Matusis:2000jf} that in noncommutative field theories it is only absent, if  supersymmetry is exact.
For softly broken supersymmetry only the IR-singular (pole) contribution to $\Pi_2$ vanishes,
but a constant term
\begin{equation}
\label{nonsusy}
\Pi_{2}\sim\Delta M^{2}_{\textrm{SUSY}},\quad
\Delta M^2_{\rm SUSY}=\frac{1}{2}\sum_s M_s^2-\sum_f M_f^2,
\end{equation}
remains.
In \eqref{nonsusy} the sums run over \emph{all} bosons and fermions. Therefore, if we have compactified
extra dimensions, we must include the Kaluza-Klein modes, effectively multiplying the four-dimensional
$\Delta M^{2}_{\textrm{SUSY}}$ by the number of Kaluza-Klein modes.
The number of contributing Kaluza-Klein modes is, again, given roughly by $N^{\textrm{IR}}_{\textrm{KK}}$
of Eq.~\eqref{flowir}.
Hence, we find
\begin{equation}
\label{pi2}
\Pi_{2}\sim N^{\textrm{IR}}_{\textrm{KK}}(k)\Delta M^{2}_{\textrm{SUSY}}\sim
\left(\frac{M^{2}_{\textrm{NC}}}{M_{c}k}\right)^{n}\quad\textrm{for}
\quad k^{2}\ll \min(M^{2}_{\textrm{NC}},\frac{M^{4}_{\textrm{NC}}}{M^{2}_{c}}).
\end{equation}

Solving the equation of motion for a photon propagating in the 3-direction $k^{\mu}=(k^{0},0,0,k^{3})$ and with
$\theta^{13}=-\theta^{31}\neq 0$ the only nonvanishing components in the noncommutativity matrix we find
for the two polarisation states of a zero mode of an unbroken trace-U(1) (see \cite{Jaeckel:2005wt} for more details),
\begin{eqnarray}
(\Pi_{1} k^2-\Pi_{2})=0\quad\textrm{for}\quad A^{\mu}=(0,1,0,0)\\\nonumber
\Pi_{1} k^2=0\quad\textrm{for}\quad A^{\mu}=(0,0,1,0).
\end{eqnarray}
The second polarisation state has an ordinary dispersion relation of a massless particle. The first, however
has a Lorentz symmetry breaking mass
\begin{equation}
m^{2}_{\textrm{LV}}\sim \frac{\Pi_{2}}{\Pi_{1}}\sim \Delta M^{2}_{\textrm{SUSY}},
\end{equation}
which is roughly constant although both $\Pi_{1}$ and $\Pi_{2}$ scale with a power law. Yet, these power laws
cancel since they are the same for $\Pi_{1}$ and $\Pi_{2}$.

\section{Weaker constraints from power law running}\label{bounds}

We found in Sect. \ref{massec} that the Lorentz violating mass term for the trace-U(1) factors
remains roughly constant. Hence trace-U(1)'s are unsuitable as photon candidates. With a similar argument
(see \cite{Jaeckel:2005wt} for more details) one finds that this also holds for mixtures of trace and traceless parts.
Therefore a suitable photon candidate must be constructed (as in four dimensions)
from an unbroken combination of traceless generators. In \cite{Jaeckel:2005wt}
we found that such a combination can only exist together with
additional unbroken U(1)'s which have nonvanishing trace.
Here the results of Sect.~\ref{powerlaw} help us, since they allow for a fast decoupling
of trace-U(1) degrees of freedom. This is in contrast to the four-dimensional case, where the (only)
logarithmic decoupling necessitated incredibly large noncommutativity scales
$M_{\textrm{NC}}\gg M_{\textrm{P}}$.
With additional (compactified) space dimensions we have power law running according to \eqref{flowir}.
This decouples the unwanted trace-U(1)'s much faster in the IR thereby weakening the constraints dramatically.

Let us now estimate the new constraints obtained from power law running.
As already mentioned earlier, current experiments probe the regime well below $M_{c}$.
To apply Eq. \eqref{resultu1} we also need $k\ll k_{s}$,
\begin{equation}
k_{s}=\frac{M^{2}_{\textrm{NC}}}{M_{c}}.
\end{equation}
This is also assured, since the discussion of Sect. \ref{simple}
shows that for $k\sim k_{s}$ the trace-U(1) and the SU($N$) have gauge couplings which are of the same order.
(Until $k\sim M_{\textrm{NC}}$ both gauge couplings are approximately equal and
power law running sets in only below $k_{s}$.)

\begin{figure}
\begin{center}
\scalebox{0.90}[0.90]{
\begin{picture}(190,180)(40,0)
\includegraphics[width=9.5cm]{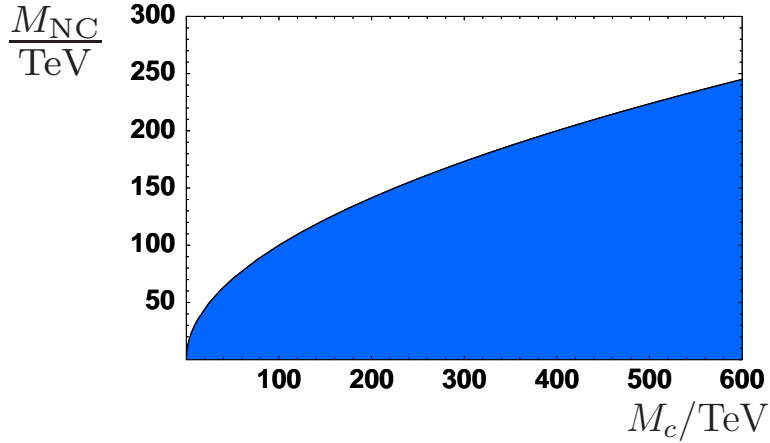}
\Text(-30,-10)[c]{\scalebox{1.4}[1.4]{$M_{c}/\textrm{TeV}$}}
\Text(-300,150)[c]{\scalebox{2.0}[2.0]{$\frac{M_{\textrm{NC}}}{\textrm{TeV}}$}}
\end{picture}
}
\end{center}
\caption{Excluded regions in the $(M_{c},M_{\textrm{NC}})$-plane
(in TeV). The blue region is excluded because the
trace-U(1) still has nonnegligible coupling.
We have chosen
$\epsilon=0.05$, $C_{1}b^{\textrm{np}}_{0}=0.1$,
$g^{2}_{\textrm{SU}(N)}(k_{0})=0.2$, $k_{0}=0.1\,\textrm{TeV}$, and
$n=1$.} \label{exclusion}
\end{figure}

Neglecting the slow logarithmic running of the SU($N$) couplings we find from Eqs. \eqref{resultu1} and \eqref{resultsun},
\begin{eqnarray}
\frac{g^{2}_{\textrm{U}(1)}}{g^{2}_{\textrm{SU}(N)}}
\!\!&\approx&\!\! \frac{n}{C^{\textrm{IR}}_{n}b^{\textrm{np}}_{0}}\frac{1}{g^{2}_{\textrm{SU}(N)}(k_{s})}
\left(\frac{k}{k_{s}}\right)^{n}
= D k^{n}\left(\frac{M_{c}}{M^{2}_{\textrm{NC}}}\right)^{n} \quad\textrm{for} \quad k\ll k_{s}\\\nonumber
D\!\!&=&\!\!\frac{n}{C^{\textrm{IR}}_{n}b^{\textrm{np}}_{0}}
\frac{1}{g^{2}_{\textrm{SU}(N)}(k_{s})}\sim \frac{(4\pi)^{2}}{4 N g^{2}_{\textrm{SU}(N)}},
\end{eqnarray}
where the $\sim$ in the second line holds for a pure noncommutative U($N$).
To have
\begin{equation}
\label{bound}
\frac{g^{2}_{\textrm{U}(1)}(k_{0})}{g^{2}_{\textrm{SU}(N)}(k_{0})}<\epsilon
\end{equation}
we need
\begin{equation}
\label{const1}
\frac{M^{2}_{\textrm{NC}}}{M_{c}}>k_{0}\left(\frac{D}{\epsilon}\right)^{\frac{1}{n}}.
\end{equation}
As an illustration we have plotted the excluded region in Fig. \ref{exclusion}.
This shows that when we allow for a 5\,\% uncertainty in the electromagnetic coupling at 100~GeV, 
the allowed region of $M_{\rm NC}$ starts already at a few TeV, depending on the compactification scale. 

\section{Summary and conclusions}

We have demonstrated that in a noncommutative U($N$) gauge theory with compact extra dimensions,
the ultraviolet/infrared mixing effects lead to a fast power-like decoupling of the
trace-U(1) degrees of freedom.

The Lorentz violating mass
term of the trace-U(1)
degrees of freedom remains roughly the same as in four dimensions.
Therefore, the trace-U(1) or mixtures of trace and traceless U(1) degrees of freedom are
unacceptable
as photon candidates. However, the fast decoupling of trace-U(1) degrees of freedom allows to
effectively hide them with only relatively mild restrictions on $M_{\textrm{NC}}$.

This allows a situation similar to the one considered
in \cite{Khoze:2004zc} where a gauge group
$\textrm{U}(4)\times\textrm{U}(3)\times \textrm{U}(2)$ was broken down to
$\textrm{SU}(3)\times\textrm{SU}(2)\times \textrm{U}(1)^{4}$, where three of the four U(1)'s have non vanishing trace.
In four dimensions these trace-U(1) would have led to observable Lorentz symmetry violation. In a
scenario with compact extra dimensions, however, the additional trace-U(1) groups may 
have very small couplings to the Standard Model matter due to the power-law decoupling in the infrared.

\bigskip
\bigskip

\centerline{\bf Acknowledgements}

We are grateful to Chong-Sun Chu and Dumitru Ghilencea for discussions.
SAA and VVK thank CERN Theory Group for hospitality in summer 2005.
VVK acknowledges the support of PPARC through a Senior Fellowship. 

\begin{appendix}
\section{Infrared running from UV/IR mixing}\label{appirrunning}

In this appendix we concentrate on the case of a noncommutative theory
with commutative compact extra dimensions, i.e. $\theta^{\mu\nu}$ is
nonvanishing in 4 dimensions and
\begin{equation}
\label{fourdcase2}
\theta^{a b}=0,\,\,\,
\theta^{\mu b}=0,\,\,\,
\textrm{for all}\,\,\, a,\, b=4,\ldots,3+n.
\end{equation}
The more complicated case of noncommutativity extending also to compact
extra dimensions will be discussed in Appendix \ref{detailed}.

Since $\theta^{a b}=0$ compact extra dimensions appear in the analysis only
via additional summations over the tower of Kaluza-Klein modes.
Hence we can follow the four-dimensional approach of \cite{Khoze:2000sy,Hollowood:2001ng}
in a noncommutative gauge theory with generic massive fields, and sum over Kaluza-Klein modes
in the end.

The running (Wilsonian) gauge coupling is defined via
\begin{equation}
\label{defcoupling2}
\left(\frac{1}{g^{2}}\right)^{AB}=\left(\frac{1}{g^{2}_{0}}\right)^{AB}+\Pi^{AB}_{1}(k),
\end{equation}
where $\Pi_{1}$ is part of the polarisation tensor given in Eq. \eqref{poltensor}.
UV/IR mixing appears only in the trace-U(1) part of the gauge coupling, corresponding to the $(0,0)$ component
in our conventions. The SU($N$) part is unaffected by the UV/IR effects and
 behaves like in a usual commutative gauge theory. Therefore, we
will concern ourselves only with the trace-U(1), i.e. the $(0,0)$ component and drop the index.
In noncommutative field theories loop integrals contain additional factors of $\exp(i\frac{p\tilde{k}}{2})$ where
$p$ is the loop momentum and $k$ an external momentum. Using trigonometric relations one can combine these
factors into parts proportional to unity, the so called planar parts, and parts proportional to $\exp(ip\tilde{k})$,
the non-planar parts. For the $\Pi_{1}$ part we explicitly write
\begin{equation}
\label{plnpldec}
\Pi_{1}=\Pi^{\textrm{planar}}_{1}+\Pi^{\textrm{np}}_{1}.
\end{equation}

At one loop and using dimensional regularisation one finds \cite{Hollowood:2001ng,Khoze:2000sy}
\begin{eqnarray}
\label{planarsusy2}
&&\Pi_{1\,\textrm{planar}} (k^2) =
-{4 \over (4\pi )^2 }\bigg( \sum_{j, {\bf r}} \alpha_{j}  C({\bf r})
\bigg[2C_j+\frac{8}{9}d_j
\\\nonumber
&&\quad\quad\quad\quad\quad\quad\quad\quad\quad\quad\quad\quad\quad+
\int_{0}^{1} dx \left(C_j-(1-2x)^2d_j\right)\ \log {A(k^2 , x,m^2_{j, {\bf r}}) \over \Lambda^2} \bigg]\bigg),
\end{eqnarray}
where $m_{j, {\bf r}}$ is the mass of a spin $j$ particle belonging to the representation $\bf r$ of the gauge group,
\begin{equation}
A(k^2,x,m^2_{j, {\bf r}})=k^2 x(1-x)+m^2_{j, {\bf r}},
\end{equation}
and
$\Lambda$  appears via dimensional transmutation similar to $\Lambda_{\overline{\textrm{MS}}}$ in QCD.
We have chosen the renormalisation scheme, i.e. the finite constants, such that $\Pi_{1\,\textrm{planar}}$ vanishes
at $k=\Lambda$.
\begin{table}[!t]
\begin{center}
\begin{tabular}{|c|c|c|c|c|}
\hline  j=& scalar  & Weyl fermion & gauge boson  & ghost \\
\hline $\alpha_{j}$ & -1 & $\frac{1}{2}$ & $-\frac{1}{2}$ &1  \\
\hline  $C_j$ & 0 &  $\frac{1}{2}$& 2 & 0 \\
\hline  $d_j$ & 1 & 2 & 4 & 1 \\
\hline
\end{tabular}
\end{center}
\caption{Coefficients appearing in the evaluation of the loop diagrams.}
\label{coefficients}
\end{table}

For the trace-U(1) part the nonplanar parts do not vanish and we find
\begin{equation}
\Pi_{1}^{\rm np} = { 1\over 2 k^2}\left(\hat{\Pi} - \tilde{\Pi}\right) ,
\end{equation}
with
\begin{eqnarray}
\hat{\Pi} &=& 2{C({\bf G}) \over (4\pi)^2}\sum_{j} \alpha_{j} \left\{
{8 d_j \over \tilde{k}^2} - k^2\left[ 12C_j - d_j\right]
\int_{0}^{1}dx \
K_{0} (\sqrt{A} |\tilde{k}|)\right\} ,
\\
\tilde{\Pi}& =& {8C({\bf G})\over (4\pi)^2}\sum_{j} \alpha_{j} \left\{
{ d_j \over \tilde{k}^2}-  \left(C_j k^2 -
d_j
{\partial^2 \over \partial^2 |\tilde{k}| }  \right)
 \int_{0}^{1}dx \
K_0 (\sqrt{A} |\tilde{k}|)\right\} ,
\end{eqnarray}
where $C({\bf G})=N$ is the Casimir operator in the adjoint representation.

In a supersymmetric theory the numbers of bosonic and fermionic degrees of freedom match, and
the above expressions can
be considerably simplified by using the relation
\begin{equation}
\label{susyreln}
\sum_{j} \alpha_{j} d_{j}=0.
\end{equation}
It follows that
\begin{eqnarray}
\label{definitions0}
\Pi^{\textrm{planar}}_{1}(k)\!\!&=&\!\!
-{4 \over (4\pi )^2 }\bigg( \sum_{j, {\bf r}} \alpha_{j}  C({\bf r})C_j
\bigg[2+\int_{0}^{1} dx \ \log {A(k^2 , x,m^2_{j, {\bf r}}) \over \Lambda^2} \bigg]\bigg)
\\
\Pi^{\textrm{np}}_{1}(k) &=& -{8C({\bf G}) \over (4\pi)^2} \sum_{j}\alpha_{j}C_{j}
\int_{0}^{1}dx \
K_{0} (\sqrt{A} |\tilde{k}|)
\end{eqnarray}
where in $\Pi^{\textrm{np}}_{1}$ only adjoint representations contribute.

We are now ready to take into account effects of the tower of Kaluza-Klein modes.
In a supersymmetric theory all members of a supermultiplet have the same mass $m_n$
for each Kaluza-Klein mode $n.$
Thus the contribution of the each non-zero Kaluza-Klein mode into the polarisation tensor
(and the running of the coupling) is obtained from the equations above by setting all masses
equal to $m_n.$
We find,
\be
\label{resultKK}
\Pi^{\textrm{planar}}_{1}(k;m_n) = \, \frac{1}{2}b^{\textrm{p}}_{0} F^{\textrm{p}}(k,m_n)\ , \qquad
\Pi^{\textrm{np}}_{1}(k;m_n) = \, b^{\textrm{np}}_{0}F^{\textrm{np}}(\tilde{k},m_n) \ ,
\ee
where we have defined
\begin{eqnarray}
\label{definitionsa}
b^{\textrm{p}}_{0} &=& -{8 \over (4\pi )^2 } \sum_{j, {\bf r}} \alpha_{j}  C({\bf r})C_j >0
\\
\label{definitionsb}
b^{\textrm{np}}_{0} &=& -{8C({\bf G}) \over (4\pi)^2} \sum_{j}\alpha_{j}C_{j} >0
\\
\label{definitionsc}
F^{\textrm{p}}(k,m_n) &=& 2+\int_{0}^{1} dx \ \log {A(k^2 , x,m_n^2) \over \Lambda^2}
\\
\label{definitionsd}
F^{\textrm{np}}(\tilde{k},m_n) &=& \int_{0}^{1}dx \ K_{0} (\sqrt{A} |\tilde{k}|)
\end{eqnarray}
As we have indicated above, both
the ``planar'' and ``non-planar'' $\beta$-function coefficients $b^{\textrm{p}}_{0}$
and $b^{\textrm{np}}_{0}$ are positive in asymptotically free theories.
(For an example, each non-zero KK mode of a simple ${\cal N}=2$ SYM pure gauge theory
contributes $b^{\textrm{p}}_{0} = N /(4\pi^2)$ and $b^{\textrm{np}}_{0} = N /(4\pi^2).$)
The planar coefficient $b^{\textrm{p}}_{0}$ receives contributions from all allowed representations:
adjoint and fundamental, while the non-planar one, $b^{\textrm{np}}_{0}$, receives
contributions only from adjoint fields.

The task is now boiled down to deriving approximate formulas for $F^{\textrm{p}}$ and $F^{\textrm{np}}$.
We note that $F^{\textrm{p}}$
depends on the scale $\Lambda$ but not on $\theta$ as appropriate for a planar part, while
$F^{\textrm{np}}$ depends on $\theta$ ($|\tilde{k}|=\theta |k|$, see Eqs. \eqref{ktilde} and \eqref{absktilde})
but not on $\Lambda$. The latter is a consequence of the
fact that the non-planar parts are finite.

\begin{figure}
\begin{center}
\subfigure{\scalebox{0.7}[0.7]{
\begin{picture}(190,180)(40,0)
\includegraphics[width=9.5cm]{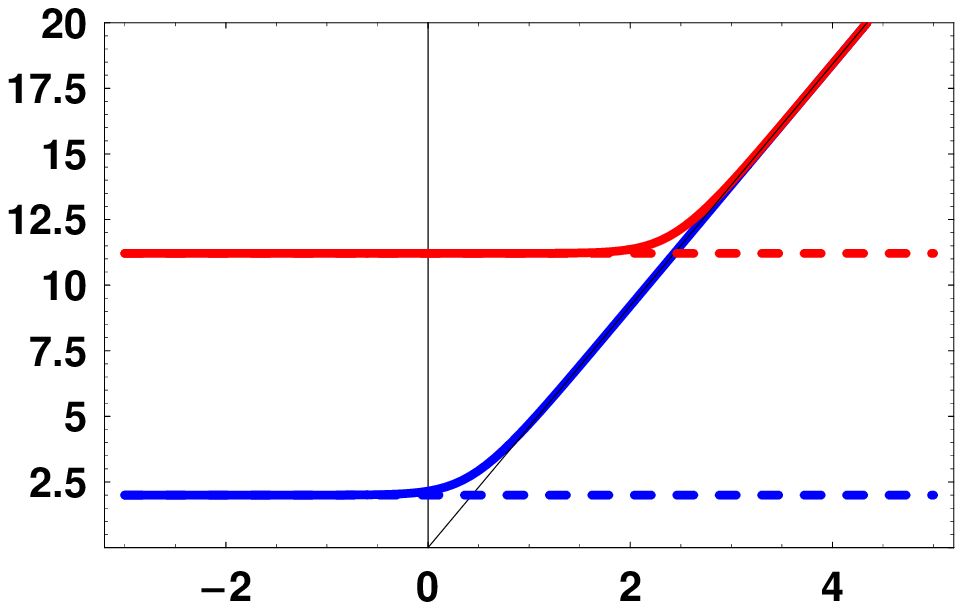}
\Text(-30,-10)[c]{\scalebox{1.4}[1.4]{$\log_{10}(k)$}}
\Text(-280,150)[c]{\scalebox{1.6}[1.6]{$F^{\textrm{p}}$}}
\end{picture}
}}
\hspace{3cm}
\subfigure{\scalebox{0.7}[0.7]{
\begin{picture}(190,180)(40,0)
\includegraphics[width=9.5cm]{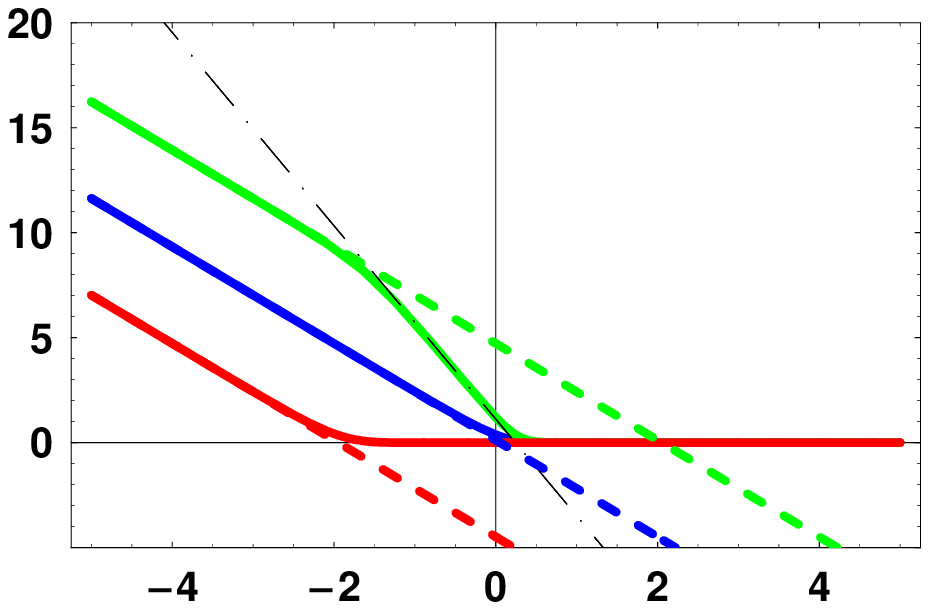}
\Text(-30,-10)[c]{\scalebox{1.4}[1.4]{$\log_{10}(k)$}}
\Text(-280,150)[c]{\scalebox{1.6}[1.6]{$F^{\textrm{np}}$}}
\end{picture}
}}
\end{center}
\caption{Functions $F^{\textrm{p}}$ (left panel) and $F^{\textrm{np}}$ (right panel)
appearing in the 1-loop calculation of
the gauge coupling. In the left panel the blue and red solid lines are
for $m=1\,\Lambda$ and $m=100\,\Lambda$, respectively. The
dashed lines give the leading order result for $k\ll m$ while the thin black line depicts the
leading order result for $k\gg m$. In the right panel the solid
lines are for $m=0.01\,\theta^{-\frac{1}{2}}$ (green), $m=\theta^{-\frac{1}{2}}$ (blue)
and $m=100\,\theta^{-\frac{1}{2}}$ (red). The dashed lines depict the approximation for
$k\gg m$ and $k\gg \theta^{-\frac{1}{2}}$. The black dot dashed line approximates in the intermediate
range $m\ll k \ll \theta^{-\frac{1}{2}}$.}
\label{functions}
\end{figure}

To illustrate the behavior of $F^{\textrm{p}}$ and $F^{\textrm{np}}$ we have plotted them in Fig. \ref{functions}.
{}From the figure it is already clear that the lowest order approximations will be either
constants or linear functions of $\log(k)$, depending on the regime we are looking at.

An important point is that $F^{\textrm{p}}$ determines the KK-mode contribution to the running
of the $1/g^2$ coupling in the UV regime (large momenta), while
$F^{\textrm{np}}$ contributes to the running only in the IR region (small momenta).
Also the slopes of $F^{\textrm{p}}$ and $F^{\textrm{np}}$ are opposite
which leads to asymptotic freedom in the UV and simultaneously to a decoupling
of the trace-U(1) factor in the IR\footnote{We note that
$F^{\textrm{np}}$ is entirely due to UV/IR mixing, and it is non-zero
only for the trace-U(1) coupling, while the planar term $F^{\textrm{p}}$ is present for
all U($N$) degrees of freedom.}.
The change of slope from UV to IR leads to the opposite sign in the right hand sides of the
flow equations for the UV running \eqref{flow}
and the IR running \eqref{flowir}
of the trace-U(1) factor.

Both functions $F^{\textrm{p}}$ and $F^{\textrm{np}}$ can be well approximated analytically.
Let us start with $F^{\textrm{p}}$. Making a series expansion of the integrand in \eqref{definitionsc}
and then integrating over $x$ we find,
\begin{eqnarray}
\label{fpapprox}
F^{\textrm{p}}\!\!&\approx&\!\!2+\log(m^2)\quad \textrm{for}\quad k^2\ll m^2,\\\nonumber
F^{\textrm{p}}\!\!&\approx&\!\!\log(k^2)\quad \textrm{for}\quad k^2\gg m^2.
\end{eqnarray}
Combining these two limits at their intersection we find
\begin{equation}
F^{\textrm{p}}\approx (2+\log(m^2))\Theta\left(1-e^2\frac{k^2}{m^2}\right)
+\log(k^2)\Theta\left(e^2\frac{k^2}{m^2}-1\right),
\end{equation}
where $\Theta(z)$ is the step-function.
As we can see from Fig. \ref{functions} these approximations work very well aside from some
(small) threshold effects around $k^2=m^2$.

The same procedure can be applied to $F^{\textrm{np}}$,
\begin{eqnarray}
F^{\textrm{np}}\!\!&=&\!\!0\quad\textrm{for}\quad (k^2\gg m^2\wedge k^2\gg \theta^{-1})
\\\nonumber
F^{\textrm{np}}\!\!&=&\!\!1-\gamma+\log(2)-\log(k^2\theta)\quad\textrm{for}\quad
(k^2\ll \theta^{-1} \, {\rm and}\,  k^2 \gg m^2)
\\\nonumber
F^{\textrm{np}}\!\!&=&\!\!-\gamma+\log(2)-\log(k m\theta)\quad\textrm{for}\quad
(k^2\ll \theta^{-1}  \, {\rm and}\,   k^2 \ll m^2),
\end{eqnarray}
where $\gamma$ is Euler's constant. Combining we find,
\begin{equation}
F^{\textrm{np}}=1-\gamma+\log(2)-\log(k^2\theta)\Theta\left(1-\frac{k^2\theta}{2e^{1-\gamma}}\right)
+\frac{1}{2}\log\left(\frac{k^2}{e^2 m^2}\right)\Theta\left(1-\frac{k^2}{e^2 m^2}\right).
\end{equation}

Our results for $F^{\textrm{np}}$ together with Eq.~\eqref{resultKK} determine
the contributions of each non-zero Kaluza-Klein mode to the running coupling $1/g^2$
and the corresponding flow equation \eqref{flowir} in the infrared.
The factor $N^{\textrm{IR}}_{\textrm{KK}}(k)$ on the right hand side of \eqref{flowir} arises
from summing over all contributing Kaluza-Klein modes.

Similarly and as a bonus the asymptotic expression for
$F^{\textrm{p}}$ together with Eq.~\eqref{resultKK} also confirm
the UV flow equation \eqref{flow} for the U($N$) coupling.

\section{Detailed calculation of the running gauge coupling}\label{detailed}

In this appendix we perform a detailed calculation for the sum over the Kaluza-Klein modes.
This will result in a better determination of the prefactor of the power law and a more general applicability.
In particular,
the calculation is also valid for noncommutativity that extends to the extra dimensions.
It is convenient to stick with the dimensional regularisation, and in fact, as we shall see this makes
little difference for the IR regime.

As already mentioned earlier, we restrict ourselves to the
case that the \emph{external} momentum $k$ is purely four dimensional,
i.e. we are interested in situations where the external particles are the (approximate)
zero modes of the Kaluza-Klein spectrum.

We start from the four-dimensional one loop formula for the polarisation tensor
and re-express the decomposition \eqref{plnpldec} as follows
(cf. \cite{Alvarez-Gaume:2003,Khoze:2000sy,Hollowood:2001ng}):
\begin{eqnarray}
\label{combine}
\Pi^{\textrm{4-dim}}_{\mu\nu}(k)=\Pi^{\textrm{4-dim}}_{\mu\nu}(k,\ell=0)
-\Pi^{\textrm{4-dim}}_{\mu\nu}(k,\ell=\tilde{k})
\label{pi}
\end{eqnarray}
with
\begin{eqnarray}
\label{basic0}
\Pi^{\textrm{4-dim}}_{\mu\nu}(k,\ell)&=&C(\mathbf{G})\sum_{j} \alpha_j\int {d^4 p\over (2\pi)^4}
\left\{ d_{j}\left[{(2p+k)_{\mu} (2p+k)_{\nu} \over
(p^2+m_{j}^2)[(p+k)^2+m_{j}^2]}-{2\delta_{\mu\nu}\over
p^2+m_{j}^2}\right]\right. \nonumber \\
&& \quad\quad\quad\quad\quad\quad\quad\quad\quad\quad\,\,\,+\left. 4 C_j{k^2 \delta_{\mu\nu}-k_{\mu}k_{\nu} \over
(p^2+m_{j}^2)[(p+k)^2+m_{j}^2]}\right\} e^{ip\cdot \ell}.
\label{effect}
\end{eqnarray}
Equation \eqref{effect} is essentially the original expression for the polarisation tensor,
which after being integrated would lead to equations such as \eqref{planarsusy2}.
The coefficients $\alpha_{j},C_{j},d_{j}$ were already  given in Table \ref{coefficients} in Appendix \ref{appirrunning}.

At present we will restrict ourselves to a supersymmetric theory. In this case the first two terms drop out,
due to Eq.~\eqref{susyreln}, and we get
\begin{eqnarray}
\label{basic}
\Pi^{\textrm{4-dim}}_{\mu\nu}(k,\ell)&=&4C(\mathbf{G})\sum_{j} C_j \alpha_j\int {d^4 p\over (2\pi)^4}
 {k^2 \delta_{\mu\nu}-k_{\mu}k_{\nu} \over
(p^2+m_{j}^2)[(p+k)^2+m_{j}^2]} e^{ip\cdot \ell}.
\label{effect2}
\end{eqnarray}

Before continuing we introduce some notation.
In the following the ``inverse'' (similarly for division by a vector) of a vector is simply the vector where
we have taken the inverse of every component, i.e.
\begin{equation}
(a_{1},a_{2},\ldots,a_{N})^{-1}=(a_{1}^{-1},a_{2}^{-1},\ldots,a_{N}^{-1}).
\end{equation}
In the same spirit component-wise multiplication of vectors is understood, when the appropriate result is a vector.
To avoid confusion we denote the ordinary scalar product by a dot: $k\cdot l$. Moreover, we abbreviate,
\begin{eqnarray}
\tilde{k}^{\mu} & = & \theta^{\mu\nu}k_{\nu}\,\,\,\,\,\,\,(\mu=0\ldots,3)\\
\hat{k}^{a} & = & \theta^{a\nu}k_{\nu}\,\,\,\,\,\,\,(a=4\ldots,3+n).
\end{eqnarray}
We will denote the Kaluza-Klein mode vector by $n$, and the vector of compactification radii as $R$.
In particular, \[\left|\frac{n}{R}\right|\] is the mass of the Kaluza-Klein mode\footnote{Note
that $n$ which labels Kaluza-Klein modes should not be confused with $n$ which counts
extra dimensions $n=D-4.$ }.

To account for the Kaluza-Klein modes we introduce the following sum as a prefactor
in front of the non-planar term in \eqref{basic},
\begin{equation}
\label{prefactor}
\sum_{n\in\mathbb{Z}^{n}} e^{i\frac{n}{R}\cdot\hat{k}}.
%e^{ik.\tilde{p}}
\end{equation}
This is the obvious multi-dimensional expression restoring as it does the
$D$-dimensional Lorentz invariance of the phase factor in \eqref{effect2}.

We now use the
Poisson resummation identity
\begin{equation}
\sum_{n\in \mathbb{Z}^{n}}\delta(\hat{\pi}-n)=\sum_{m\in \mathbb{Z}^{n}}e^{2\pi im\cdot\hat{\pi}}
\end{equation}
where we introduced a {}``dummy'' $n$-momentum for the extra dimensions,
$\hat{\pi}$. The idea is to replace the KK-sum in the integral with
a loop momentum integration. Calling for convenience $L=(l,\hat{l})$
so that we can treat planar and non-planar terms at the same time,
the phase factor in the integrals can be written as
\begin{eqnarray*}
\int d^{n}\hat{\pi}\sum_{n\in \mathbb{Z}^{n}}\delta(\hat{\pi}-n)(e^{i\frac{n}{R}\cdot\hat{l}}e^{ip\cdot l}) & = &
\sum_{m\in \mathbb{Z}^{n}}\int d^{n}\hat{\pi}\, e^{2\pi im\cdot\hat{\pi}}(e^{i\frac{\hat{\pi}}{R}\cdot\hat{l}}e^{ip\cdot l}).
\end{eqnarray*}
Rescaling $\hat{p}=\hat{\mathbf{\pi}}/R$ we get a prefactor of
\begin{eqnarray}
 &  & (\prod_{i}2\pi R_{i})\,\sum_{m\in \mathbb{Z}^{n}}\int\frac{d^{n}\hat{p}}{(2\pi)^{n}}\, e^{2\pi i(mR)\cdot \hat{p}}(e^{iP\cdot L})\nonumber \\
 & = & (\prod_{i}2\pi R_{i})\,\sum_{m\in \mathbb{Z}^{n}}\int\frac{d^{n}\hat{p}}{(2\pi)^{n}}\,(e^{iP\cdot L_{m}})
\end{eqnarray}
where
\begin{equation}
L_{m}=L+(0,2\pi mR)
\end{equation}
As usual the point of this procedure is that eventually the leading
$m=0$ term dominates. For the moment however we will keep the $m$
summation. On replacing the masses in the propagators in \eqref{effect2} with the KK
masses $m_{j}=|\hat{p}|,$ the task is reduced to evaluating the following
$D=4+n$ dimensional integral in $P=(p,\hat{p})$ where $K=(k,\,0)$:
\[
\Pi_{\mu\nu}(K,L)=4\sum_{j} \alpha_{j}C_{j}\, C(\mathbf{G})(\prod_{i}2\pi R_{i})\,(k^{2}
\delta_{\mu\nu}-k_{\mu}k_{\nu})\,\sum_{m\in \mathbb{Z}^{n}}\int\frac{d^{D}P}{(2\pi)^{D}}\frac{e^{iP\cdot L_{m}}}{|P|^{2}|K+P|^{2}}.\]
Using the identity
\[
\frac{1}{A_{1}A_{2}}=-\int_{0}^{1}dx\,\int_{0}^{\infty}dt\, te^{it\,(xA_{1}+(1-x)A_{2})}\]
for the propagators, integrating over $P'=P+(1-x)K$,
and using $d^{D}P=\frac{2\pi^{\frac{D}{2}}}{\Gamma(\frac{D}{2})}|P|^{D-1}d|P|$
we find for $L_{m}\neq 0$, 
\begin{eqnarray}
\int\frac{d^{D}P}{(2\pi)^{D}}\frac{e^{iP\cdot L_{m}}}{|P|^{2}|K+P|^{2}} & = &
(4\pi)^{-\frac{D}{2}}\int_{0}^{1}dx\, e^{i(1-x)K\cdot L_{m}} \times\nonumber \\
 &  & \,\,\,\,\,\,\,\,\,\,\,\,\,\,\,\int_{0}^{\infty}dt\, t^{(1-\frac{D}{2})}
 \exp(-t\, x(1-x)K^{2}-\frac{L_{m}^{2}}{4t})\\
 & = & \frac{1}{2}(2\pi)^{-\frac{D}{2}}\int_{0}^{1}dx\, e^{i(1-x)K\cdot L_{m}}\,
 |L_{m}|^{-n}(|L_{m}|\Delta)^{\frac{n}{2}}K_{\frac{n}{2}}\left(|L_{m}|\Delta\right), \nonumber
 \end{eqnarray}
where
\begin{equation}
\Delta=|K|\sqrt{x(1-x)}.
\end{equation}
Note that the length $|L_{m}|$ is acting as an effective UV cut-off
on the Schwinger parameter $t$. When $L_{m}=0$, (the $m=0$ planar
term) we find
\begin{equation}
\int\frac{d^{D}P}{(2\pi)^{D}}\frac{1}{|P|^{2}|K+P|^{2}}=(4\pi)^{-\frac{D}{2}}\int_{0}^{1}dx\Delta^{n}\Gamma(-\frac{n}{2}).
\end{equation}
 When $D=4$ these expressions agree with Ref.~\cite{Alvarez-Gaume:2003}.

Next we consider the summation over $m$ in order to justify throwing
away everything except the first term. First we should point out that
the summation is formally divergent. We will return to this fact shortly,
but for the moment we shall neglect it and concentrate on just the leading
terms. The asymptotic behaviour of the Bessel function in the two
limits is \begin{eqnarray}
K_{\nu}(z) & = & \sqrt{\frac{\pi}{2z}}e^{-z}\,\,\,\,\,\,\,\, z\rightarrow\infty , \nonumber \\
K_{\nu}(z) & = & \frac{\Gamma(\nu)}{2}\left(\frac{z}{2}\right)^{-\nu}\,\,\,\,\,\,\,\, z\rightarrow0 .\end{eqnarray}
The control parameter is $z=|L_{m}|\Delta\sim|L_{m}||K|$. Hence in
the UV regime (defined as $|K|\gg M_{\rm NC}$) as long as $2\pi R_{i}|K|\gg1$
the leading term in the $m$ summation dominates exponentially. This
is of course true by assumption since $|K|\gg1/R_{i}\sim M_{c}$ is
also the condition that $M_{\rm NC}$ is well above the mass of the lightest
KK states, $M_{c}$. For the IR regime ($|K|\ll M_{\rm NC}$) recall that
$L_{m}=(0,\tilde{k}+2\pi Rm)\sim(0,|k|/M_{\rm NC}^{2}+2\pi R_{i}m_{i})$.
Hence if\begin{equation}
|k|\ll2\pi\frac{M_{\rm NC}}{M_{c}}M_{\rm NC}\end{equation}
then the leading $m=0$ term is dominant, the others still being suppressed.
Again this condition is always true since $M_{\rm NC}\gg M_{c}$ and $|k|\ll M_{\rm NC}$
in this regime, however the suppression is now only polynomial when
$z\sim|2\pi R||K|\ll1$ or in other words when $|k|$ is less than
the lightest KK states. But since the Bessel functions go as $(|L_{m}||K|)^{-\frac{n}{2}}$
the leading term has an IR pole $(|k|/M_{\rm NC})^{-n}$, with the subleading
terms in the $m$ summation contributing only finite corrections.
The problem is exactly equivalent to that of finding the Green function
on a $n$-torus: close to the $\delta$-function source one recovers
the usual non-compact $r^{-n}$ result one would expect. Indeed, returning
to the question of convergence of the whole series, the $m$ summation
corresponds to the lattice of $\delta$-function images on the covering
space, and is formally infinite. By Poisson resummation the sum is
equivalent to a generalized Riemann $\zeta$-function and indeed one
can use $\zeta$-function regularisation to regulate it: for more
details, see Ref.~\cite{elizalde}. The net result of such a procedure
is that the leading $k\rightarrow0$ pole dominates as one would expect
and the two point function is well approximated by the first term
\[
\Pi_{\mu\nu}(K,L)=4\sum_{j}\alpha_{j}C_{j}\,C(\mathbf{G})(\prod_{i}2\pi R_{i})\,(k^{2}\delta_{\mu\nu}-k_{\mu}k_{\nu})\,
\int\frac{d^{D}P}{(2\pi)^{D}}\frac{e^{iP\cdot L}}{|P|^{2}|K+P|^{2}}.\]
The planar term gives the usual UV divergence and needs to be regularised.
This has been the subject of quite a few papers (for example
refs.\cite{ghilencea,oliver})
but for completeness we will do a naive dimensional regularisation
here. For ease of notation define $\eta\approx\epsilon+n/2$ with
$\eta$ being the continuously varying dimensionality, $\epsilon\rightarrow0$
and $n$ being integer.
We then get
\begin{eqnarray}
\Pi_{\mu\nu}(K,0) \!\!\! & = &\!\!\! 4\sum_{j}\alpha_{j}C_{j}\,C(\mathbf{G})(\prod_{i}2\pi R_{i})\,
(k^{2}\delta_{\mu\nu}-k_{\mu}k_{\nu})\,(4\pi)^{-\frac{D}{2}}\int_{0}^{1}dx
\Delta^{2\eta}\Gamma(-\eta)\\
 \nonumber
 & = &\!\!\! -4\sum_{j}\alpha_{j}C_{j}\,C(\mathbf{G)}(\prod_{i}2\pi R_{i})\,(k^{2}\delta_{\mu\nu}-k_{\mu}k_{\nu})\,
 (4\pi)^{-(2+\eta)}\,|k|^{2\eta}\frac{\pi\Gamma(1+\eta)}{\sin(\pi\eta)\Gamma(2+2\eta)}
 \end{eqnarray}
which gives (for even $d$)
\begin{eqnarray}
(-1)^{\frac{n}{2}}(4\pi)^{-\frac{D}{2}}|k|^{n}\frac{\Gamma(1+\frac{n}{2})}{\Gamma(2+n)}
& \times & \left(\frac{1}{\epsilon}+\psi(1+\frac{n}{2})+2\psi(2+n)+\log(4\pi|k|^{2})\right)\\
\nonumber
\end{eqnarray}
For $n$ odd the term converges. The nonplanar term can be evaluated
in the small $|k| |\tilde{k}|=k^{2}/M_{\rm NC}^{2}$ limit as
\begin{equation}
\frac{1}{(4\pi)^{2}}\pi^{-\frac{n}{2}}\Gamma\left(\frac{n}{2}\right)|\tilde{k}|^{n}
\end{equation}
This is the IR power law running advertised in the main body of the paper.

What the rather broadbrush approach presented here misses are the
contributions in the sum where some of the $m$- modes are zero and
others infinite (when $D\geq6$). This is a different kind of UV/IR
effect, that leads to higher dimensional operators in the effective
Lagrangian. Since this is a technicality at the UV end of things and
relevant only for $D\geq6$, we simply refer the reader to Ref.~\cite{ghilencea}
for more details.

\end{appendix}

\end{document}